\documentclass[prl,twocolumn,tightenlines,floatfix,showpacs,superscriptaddress]{revtex4}

\usepackage{graphicx}

\begin{document}

\bibliographystyle{prsty} 

\title{Dynamical Nuclear Polarization by Electrical Spin Injection in Ferromagnet-Semiconductor Heterostructures}

\author{J. Strand}
\affiliation{School of Physics and Astronomy}
\author{B. D. Schultz}
\affiliation{Department of Chemical Engineering and Materials
Science, \\University of Minnesota, Minneapolis, MN 55455}
\author{A. F. Isakovic}
\affiliation{School of Physics and Astronomy}
\author{C. J. Palmstr{\o}m}
\affiliation{Department of Chemical Engineering and Materials
Science, \\University of Minnesota, Minneapolis, MN 55455}
\author{P. A. Crowell}\email{crowell@physics.umn.edu}
\affiliation{School of Physics and Astronomy}

\begin{abstract}
Electrical spin injection from Fe into Al$_x$Ga$_{1-x}$As quantum well
heterostructures is demonstrated in small ($< 500$~Oe) in-plane
magnetic fields.   The measurement is sensitive only
to the component of the spin that precesses about the internal
magnetic field in the semiconductor.  This field is much larger
than the applied field and depends strongly on the injection
current density.  Details of the observed hysteresis in the spin
injection signal are reproduced in a model that incorporates the
magnetocrystalline anisotropy of the epitaxial Fe film, spin
relaxation in the semiconductor, and the dynamical polarization of
nuclei by the injected spins.
\end{abstract}


\maketitle


The injection of spin from a conventional ferromagnetic metal into
a semiconductor is a prerequisite for several proposed
magneto-electronic devices \cite{Datta:Das:SpinTransistor}.
Although spin transport
across the ferromagnet-semiconductor (FM-S) interface has recently
been demonstrated \cite{Hammar:Johnson:1st-FM-SC_spinInjection,Ploog:Zhu:InGaAsPRL,
Jonker:Hanbicki:AlGaAsAPL, IMEC:Motsnyi:OHE}, most injection
experiments on metallic FM-S systems have required relatively
large magnetic fields, in excess of several
kilogauss, to produce a spin component {\it perpendicular} to the FM-S
interface.  The most useful properties of typical ferromagnetic thin films,
however, such as low-field switching and hysteresis, can be
exploited only by coupling to the {\it in-plane} component of the
magnetization \cite{OhnoY:Awschalom:FirstGaMaAs_spinLED}. In the
case of metallic FM-S structures, in-plane coupling has been
observed only as a small change in transport
properties \cite{Hammar:Johnson:1st-FM-SC_spinInjection} or using
optically pumped carriers
\cite{Kawakami:Awschalom:FM_imprinting_in_GaAs,
Epstein:Awschalom:FPP_in_n-GaAs}.

In this Letter we report a demonstration of electrical spin
injection in FM-S heterostructures using small ($< 500$~Oe)
in-plane magnetic fields. We
measure only the component of the spin that precesses after
injection into the semiconductor using electroluminescence polarization
(ELP) as a detection technique
\cite{Fiederling:Molenkamp:FirstBeMnZnSe_spinLED,
OhnoY:Awschalom:FirstGaMaAs_spinLED}. The effective
magnetic field inducing the precession depends strongly on the
electrical bias conditions and is dramatically enhanced at the
highest injection current densities. The origin of the hysteresis
in the spin polarization signal is magnetization reversal in the
ferromagnet, but the magnitude and shape of the observed loops
depend on the effective field in the semiconductor. Modeling based
on the results of optical pumping experiments demonstrates that
the origin of the large effective field is dynamical nuclear
polarization due to the spin-polarized current injected from the
ferromagnet \cite{Johnson:ElectricalDNP-proposal}.  This approach
to dynamical nuclear polarization
in semiconductors is a simple alternative to the use of
optical pumping or high magnetic fields as sources of
spin-polarized electrons \cite{Smet:GateControl,Barrett:QWNMR}.

\begin{figure}
    \includegraphics{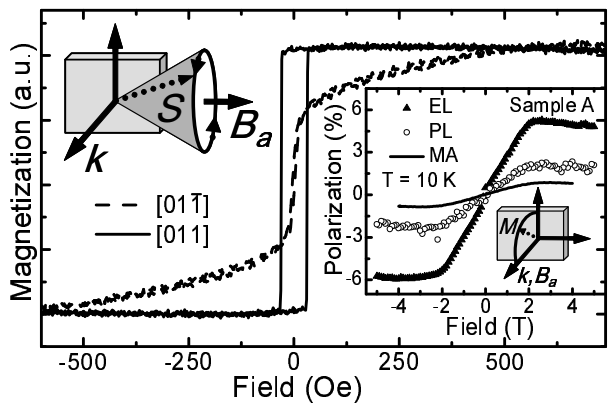}
    \caption{In-plane magnetization  of the epitaxial Fe film (sample A) parallel
to fields applied along
[011] and $[01\bar 1]$.
Left inset: schematic of Voigt geometry
experimental setup. The direction of light propagation
$\hat{\bf{k}}$ is along the sample normal, ${\bf B}_a$ is the
applied field, and $\bf{S}$ is the precessing spin in the
semiconductor. Right inset: Faraday geometry data and experimental
setup. $\bf{M}$ is the magnetization of the Fe film.  The three
curves show electroluminescence (EL), photoluminescence (PL), and
magnetoabsorption (MA) polarization as a function of the magnetic
field ${{\bf B}_a}$ applied along $\hat{\bf z}$.}
    \label{fig:Magnetization}
\end{figure}

We report results from two heterostructures with different quantum
well (QW) spin detectors.  The samples are grown by molecular beam
epitaxy on p$^+$ GaAs (100) substrates and consist of
p-Al$_x$Ga$_{1-x}$As/QW/n-Al$_x$Ga$_{1-x}$As/Fe/Al
\cite{Isakovic:SpinInjection}. Intrinsic setback layers are placed
on both sides of the QW, and the resulting p-i-n structure forms a
light-emitting diode (LED). The 50~{\AA} thick iron and the
25~{\AA} Al layers are grown {\it in situ} at $\approx 0^\circ$~C.
A Si $\delta$-doped layer ($\delta$ = $2\times
10^{13}$~atoms/cm$^{2}$ for sample A and $3\times
10^{13}$~atoms/cm$^{2}$ for sample B) is grown 25~{\AA} below the
Fe-semiconductor interface in order to form a tunneling contact.
For sample A, $x=0.2$ and the QW is 100~{\AA}
Al$_{0.1}$Ga$_{0.9}$As with p-doped barriers.  For sample B,
$x=0.1$ and the QW is 100~{\AA} GaAs with intrinsic barriers. The
important difference between the samples is the effective
$g$-factor $g^*$ for electrons in the QW.  For sample B, $g^*
\approx -0.4$, while sample A was designed to have an electron
$g$-factor close to zero \cite{Chadi:AlGaAs_g-factor}. The
magnetocrystalline anisotropy of the epitaxial Fe film results in
easy and hard axes along [011] and [01$\bar{1}$], as shown in
Fig.~\ref{fig:Magnetization}.

The samples are processed into 200~$\mu$m wide mesas. The
ferromagnetic contact is 80~$\mu$m wide, and a 40~$\mu$m wide gold
contact is evaporated over the center of the bar, which is cleaved
into pieces 1--1.5~mm long. The devices are operated with the Schottky
contact under reverse
bias and the p-i-n LED under forward bias. Electrons tunnel from
the Fe into the n-layer and recombine in the QW with holes
supplied by the substrate. Electroluminescence (EL) is collected
along the sample normal $\hat{\bf z}$, and is dominated by
heavy-hole recombination. Under these conditions the
circular
polarization of the EL, which we will refer to as the ELP signal,
is equal to the electron spin polarization along $\hat{\bf z}$ at
the time of recombination \cite{OO:DyakonovPerel}.

In a typical ELP measurement in the Faraday geometry, $\hat{\bf k}
\parallel \hat{\bf z}
\parallel {\bf B}_a$, where $\hat {\bf k}$ is the direction of light
propagation, and ${\bf B}_a$ is the applied magnetic field
\cite{Ploog:Zhu:InGaAsPRL, Jonker:Hanbicki:AlGaAsAPL}.
The inset of Fig.~\ref{fig:Magnetization} shows the ELP signal
measured in the Faraday geometry for sample A.  The observed
signal is proportional to the magnetization of the Fe film and is
significantly larger than the background determined by either the
photoluminescence (PL) polarization measured at a bias just below
the EL threshold or the magnetoabsorption (MA) of the Fe film.
This result indicates a steady-state spin polarization in the QW
of at least 4\% for applied fields above 2.1~T, the
saturation field for Fe.

In the transverse field configuration ($\hat{\bf k} \parallel
\hat{\bf z} \perp {\bf B}_a$) shown in the left inset of
Fig.~\ref{fig:Magnetization}, the magnetocrystalline anisotropy of
the Fe film allows for a significant angle between the electron
spin ${\bf S}$ and ${\bf B}_a$. The magnetization remains in the
sample plane and ${\bf S}$ precesses out of the plane after
injection into the semiconductor.  This approach is related
philosophically to the oblique Hanle effect
\cite{OO:DyakonovPerel}, in which optically pumped spins are
created at an angle $0 < \theta < \pi/2$ with respect to the
applied field, an approach followed in the recent electrical spin
injection experiment of Motsnyi {\it et al.}
\cite{IMEC:Motsnyi:OHE}.   An important advantage of our approach
is that we detect only the precessing component of the spin,
making the measurement essentially immune to background effects.
\begin{figure}
    \includegraphics{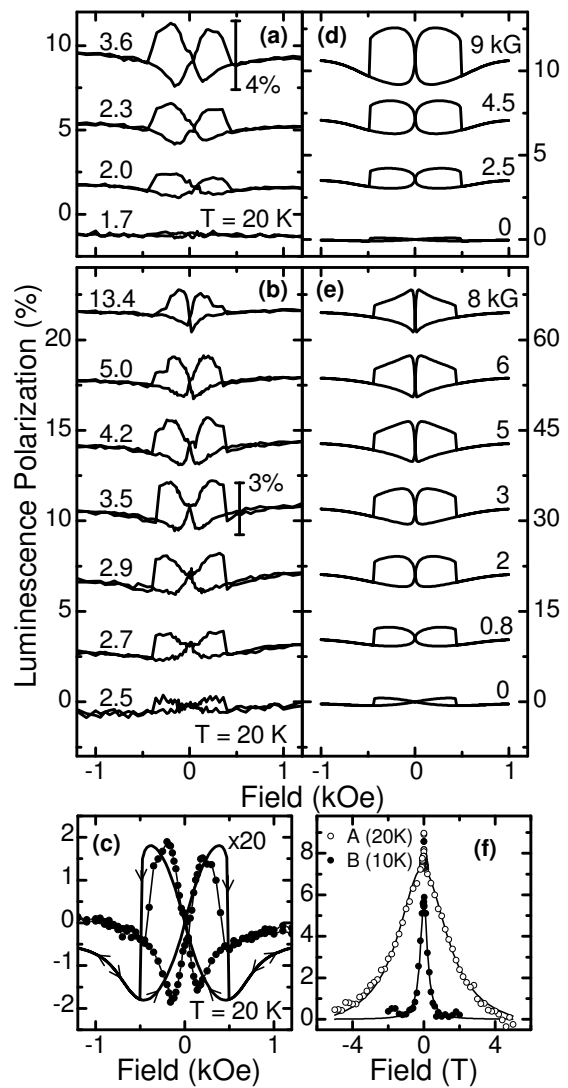}
    \caption{(a), (b) Electroluminescence polarization
(ELP) as a function of field applied at $4^\circ \pm 2^\circ$ to
[01$\bar{1}$] for samples A and B, respectively, at several
current densities. Each curve is labeled by the current density in
units of A/cm$^2$.
 (c) Sample A ELP (points) as a function of field, with a -2.15\%
 offset removed, and model ELP signal (solid curve) scaled by a
 factor of 20.  (d), (e) Model ELP signal, including
the effective nuclear field, as a function of applied field for
samples A and B, respectively.  The curves are labeled  by the
effective field $b_N$ in kG (see Eq.~\ref{eq:nuclearfield}). (f)
Hanle curves measured for samples A and B. Curves in (a), (b), (d)
and (e) are offset for clarity.}
    \label{fig:DataModel}
\end{figure}

Figures~\ref{fig:DataModel}(a) and (b) show the ELP signal as a
function of the transverse magnetic field, which is applied in the
sample plane at a small angle (4$^\circ \pm 2^\circ$) from the
[01$\bar{1}$] axis. Data are shown at several different current
densities for samples A and B.  At the lowest current densities,
just above the threshold for light emission, the ELP signal is
nearly independent of the applied magnetic field and shows no
hysteresis, although there is a polarization offset of 0 - 2.5\%.
As the current density is increased, a double loop structure
develops. In the case of sample A [Fig.~\ref{fig:DataModel}(a)],
the magnitude of the polarization signal increases with increasing
bias, reaching a maximum value of approximately 4\%, determined
from the peak-to-peak height of the loop. In the case of sample B
[Fig.~\ref{fig:DataModel}(b)], both the magnitude {\it and the
shape} of the observed loops depend on the current density. A
maximum in the signal is observed at a current density of
approximately 3.5~A/cm$^2$, above which the signal decreases and
the central dip in the loop becomes much narrower.
Photoluminescence measurements on the same structures show no
polarization signal in a transverse magnetic field.

The origin of the ELP signal observed in Fig.~\ref{fig:DataModel}
is precession of the spin injected from the ferromagnet.  The
observed polarization depends on the angle between ${\bf B}_a$ and
the magnetization ${\bf M}$ of the Fe film
as well as the precession frequency and spin relaxation
time in the QW.  Under the assumption that the magnetization
reversal in the Fe film occurs by coherent rotation
\cite{StonerWohlfarth}, the angle between ${\bf M}$ and ${\bf
B}_a$, and hence the orientation of the injected spin ${\bf S}_0$
can be determined from the
magnetization data. The spin
dynamics in the semiconductor are modeled using the formalism
developed for the optical Hanle effect \cite{  OO:DyakonovPerel}:
\begin{equation}
\frac{d{\bf S}}{dt} = \frac{{\bf S}_0}{\tau} - \frac{{\bf
S}}{\tau_s} - \frac{{\bf S}}{\tau} + {\bf \Omega}\times{\bf S},
\label{eq:spindynamics}
\end{equation}
where $\tau_s^{-1}$ is the spin relaxation rate, $\tau^{-1}$ is
the recombination rate, and ${\bf \Omega} = g^{*}\mu_B{\bf B}/
\hbar$, where $\mu_B$ is the Bohr magneton.

The steady-state spin ${\bf S}$ that is measured by ELP
can be found from Eq.~\ref{eq:spindynamics} by setting $d{\bf
S}/dt = 0$. We will be interested in an upper bound for ${\bf S}$
and will hence assume that ${\bf S}_0 = \epsilon \hat{\bf m}$,
where $\epsilon = 0.42$ is the spin polarization of Fe
\cite{Meservey:Tedrow:spin_pol_tunneling} and $\hat {\bf m}$ is a
unit vector parallel to {$\bf M$}. The parameters in
Eq.~\ref{eq:spindynamics} are thus reduced to two: the ratio
$\alpha = \tau_s/\tau$, and a characteristic field scale
$B_{1/2}=\hbar/(\mu_{B}g^{*}T_s)$, where
$T_s^{-1}=\tau_s^{-1}+\tau^{-1}$.  We determine $\alpha$ and
$B_{1/2}$ by fitting the results of Hanle effect measurements, in which electron
spins are optically injected along $\hat{\bf z}$ using circularly
polarized light tuned near the band-edge of the QW barriers.  The
circular polarization of the PL as a function of the transverse
magnetic field is shown for both samples in
Fig.~\ref{fig:DataModel}(f). Fits to the expected Lorentzian shape
\cite{OO:DyakonovPerel} are shown as the solid curves.
The dramatic difference in the widths of the two Hanle curves is
due primarily to the different electron $g$-factors in the two
quantum wells.

Once the magnetization of the ferromagnet and the Hanle curve are
known for each sample, it is then possible to model the ELP signal
without any free parameters. The result for sample A is shown as
the solid curve in Fig.~\ref{fig:DataModel}(c) along with data
obtained at a current density of 3.6 A/cm$^2$.  (A
field-independent offset has been removed from the experimental
data.)  The ELP signal vanishes at large applied fields because
the injected spin is parallel to the field, and the signal
vanishes at zero field because there is no precession.  At
intermediate fields, the magnetization is between the easy and hard
axes, and the non-zero angle between ${\bf S}_0$ and ${\bf
B}_a$ results in an ELP signal.

In spite of the general qualitative agreement in
Fig.~\ref{fig:DataModel}(c), the magnitude of the signal
determined from our model is {\it smaller} than the experimental
result by a factor of $\approx 20$, even though we have already
assumed a spin injection efficiency of unity. More significantly,
the model of Eq.~\ref{eq:spindynamics} cannot replicate the
current-driven change in the shape of the ELP loops observed for
sample B without requiring either the spin lifetimes or $g^*$ to
increase by over an order of magnitude with increasing bias.

The most probable explanation for the discrepancy between the
simple dynamical model and the experiment is that the effective
magnetic field in the semiconductor is larger than the applied
field and depends on the current density.  This hypothesis is
motivated by the observation of large internal magnetic fields in
optical pumping experiments due to dynamical nuclear polarization
(DNP) by electron spins \cite{Paget:GaAsNuclearSpinCoupling,
Flinn:Kerr:1stOptNMR_single_GaAsQW}. In our experiment, the spins
are injected electrically, and the magnitude of the nuclear
polarization is determined by the current density
\cite{Johnson:ElectricalDNP-proposal}. The total magnetic field
for electron spins is then the sum of the applied field ${\bf
B}_a$ and a nuclear field \cite{Paget:GaAsNuclearSpinCoupling}
\begin{equation}
{\bf B}_N = b_N \frac{{(\hat{\bf s} \cdot {\bf B}_a){\bf
B}_a}}{{B_{a}^2 + B_{0}^2 }},
 \label{eq:nuclearfield}
\end{equation}
where $\hat{\bf s}$ is a unit vector parallel to the injected
spin, $B_0$ is the nuclear dipolar field and $b_N$ is the maximum
nuclear field. We set $B_0$ to 2~G, a typical value for GaAs
\cite{Paget:GaAsNuclearSpinCoupling}, and allow $b_N$ to vary with
current density.  The total field ${\bf B} = {\bf B}_a + {\bf
B}_N$ is then used in Eq.~\ref{eq:spindynamics} to determine ${\bf
S}$ and the expected ELP signal.  The results of modeling with a
nuclear field,
adjusting $b_N$ as the only free parameter, are shown in
Figs.~\ref{fig:DataModel}(d) and (e). Agreement with the magnitude
of the experimental results is obtained using values of $b_N$
between 0 and 9~kG, which are indicated next to each curve. These
are comparable to the internal fields obtained in optical pumping
experiments and are well below the maximum possible internal field
of 5.3~T \cite{Paget:GaAsNuclearSpinCoupling, Salis:OptNMR-PRB}.

Including the effects of DNP also accounts for the major
differences between the ELP curves observed for samples A and B.
Although the shape of the ELP loops is nearly independent of
current density for sample A, the position of the maximum signal
observed for sample B shifts toward zero field as the current
density increases, and the jump observed at larger applied fields
disappears.  The same trend is observed in the theoretical curves.
The difference in behavior reflects the fact that the
magnetization reversal in the two samples is essentially the same,
while the widths of the Hanle curves differ by a factor of six. In
the case of sample A, the total internal field is less than the
half-width of the Hanle curve.  In this case, the average spin
precesses only a fraction of a cycle before recombination, and the
shape of the ELP loop is determined only by the rotation of the
magnetization in the ferromagnet.  In contrast, the maximum
nuclear field in sample B exceeds the half-width of the Hanle
curve, and hence spin precession in the semiconductor changes
the shape of the ELP signal as well as determining its magnitude.

Given that only one parameter was varied to obtain the theoretical
curves shown in Fig.~\ref{fig:DataModel}, the overall agreement
with the experiment provides strong support for DNP by electrical
spin injection. The most important discrepancy between the data
and the model occurs near zero field, where the suppression of the
experimental signal occurs over a much wider field range than in
the model. The narrow region observed in the model is based on the
nuclear dipole field of 2~G for bulk GaAs. However, the
suppression of DNP over a much wider field range has been observed
in optical pumping studies of ferromagnet-semiconductor interfaces
\cite{Kawakami:Awschalom:FM_imprinting_in_GaAs} and quantum dots
\cite{Gammon:QD_hyperfine_interaction}.  Furthermore, the behavior
near zero field is very sensitive to the details of the magnetic
reversal process which may not be described completely by the
simple coherent rotation model.

\begin{figure}
\includegraphics{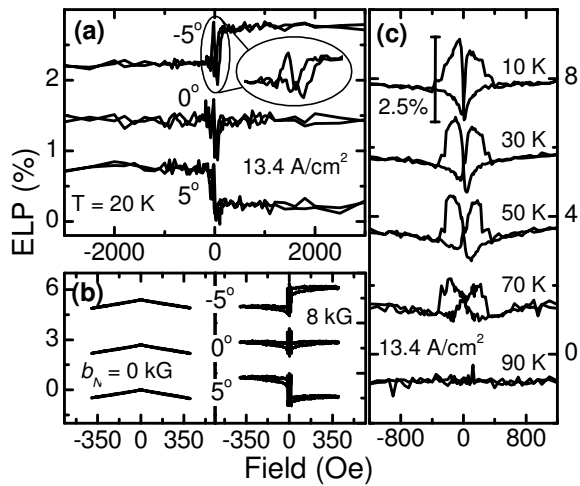}
    \caption{(a) Sample B electroluminescence polarization (ELP) data
as a function of  magnetic field for directions close to the [011]
axis.  The field is applied in the (100) plane (middle), and $\pm
5^\circ$ out of the (100) plane (top and bottom curves).  There is
an in-plane rotation of 5$^\circ$ in all three cases.  The inset
shows an expansion of the region near zero field. (b) Modeling of
results for this geometry without (left) and with (right) the
nuclear field.  (c) ELP signal for sample B as a function of
magnetic field applied along $[01{\bar 1}]$ at several different
temperatures. All curves are offset for clarity.}
    \label{fig:EasyAxis}
\end{figure}

Additional evidence for
the existence
 of a nuclear field comes from data obtained for fields applied
near the $[011]$ axis.  As can be seen from the square hysteresis
loop shown in Fig.~\ref{fig:Magnetization}, the magnetization
along this direction is essentially parallel to ${\bf B}_a$ for
any field, and in this case the torque term in
Eq.~\ref{eq:spindynamics} is zero. As a result, no precession is
expected.  This is consistent with the behavior observed in the
middle curve of Fig.~\ref{fig:EasyAxis}(a), which was obtained for
a magnetic field applied in the (100) plane at an angle of
$5^\circ$ with respect to [011].
The top and bottom curves in Fig.~\ref{fig:EasyAxis}(a) were
obtained for rotations of the field axis $\pm 5^{\circ}$ out of
the (100) plane. The curves in Fig.~\ref{fig:EasyAxis}(b) are the
results of the model of Eq.~\ref{eq:spindynamics} using only the
applied field (left) and including the nuclear field from
Eq.~\ref{eq:nuclearfield} (right). The effective field $b_N$ of
8~kG is the same as that used to fit the hard axis data at the
same current density, and  the raw magnetization data were used to
determine the direction of ${\mathbf S_0}$. As can be seen in
Figs.~\ref{fig:EasyAxis}(a) and (b), there is good agreement with
the experiment {\it only} if the nuclear field is included.  The
extreme sensitivity of the modeling results to the presence of the
nuclear field in the easy axis case follows from the dot product
$\hat{\bf s}\cdot{\bf B}_a$ in Eq.~\ref{eq:nuclearfield}.  This
compensates for the fact that the torque term in
Eq.~\ref{eq:spindynamics} nearly vanishes for small angles between
${\bf S}$ and ${\bf B}_a$. The dip in the signal at zero field
shown in the inset of Fig.~\ref{fig:EasyAxis}(a) is due to dipolar
relaxation.

The efficiency of the DNP process should decrease
with increasing temperature \cite{NucMag:Abragam}.
A strong suppression of the ELP signal with increasing
temperature is seen in both samples and is shown for sample B
in Fig.~\ref{fig:EasyAxis}(c).
The change in the shape of the ELP hysteresis loop with
temperature also mirrors the dependence on current density seen in
Fig.~\ref{fig:DataModel}.

The approach to spin injection outlined in this Letter
realizes the possibility of using conventional
ferromagnets, small magnetic fields, and DC electrical currents to
create and manipulate spin polarized carriers in a semiconductor
device.  There remain several open questions, including a
microscopic description of the DNP mechanism and its relation to the injection
process.  This will require a more detailed study of the transport properties of these
systems, with the goal of achieving a purely electronic means of spin detection.

We acknowledge L. J. Sham and C. Ciuti for helpful
discussions. This work was supported by ONR, the DARPA/ONR SPINS
program, the University of Minnesota MRSEC (NSF DMR-0212032),
and the Institute for Rock Magnetism.
A.F.I. and B.D.S thank 3M for fellowship support.


\end{document}